\documentstyle{article}\begin{document}
\title{Linearized Kompaneetz equation as a relativistic diffusion  }
\author{Z. Haba \\
Institute of Theoretical Physics, University of Wroclaw,\\ 50-204
Wroclaw, Plac Maxa Borna 9, Poland,
\\email:zhab@ift.uni.wroc.pl\\PACS numbers:05.10.Gg,05.20.Dd,98.80.Jk}\date{\today}\maketitle
\begin{abstract}
We show that Kompaneetz equation describing photon diffusion
in an environment of an electron gas, when linearized around
its equilibrium distribution, coincides with the relativistic
diffusion discussed in recent publications.
 The model of the relativistic diffusion is related to soluble models of
imaginary time quantum mechanics.  We suggest  some non-linear generalizations of the
relativistic diffusion equation and their  astrophysical applications
(in particular to the Sunyaev-Zeldovich effect).
\end{abstract}
 \section{Introduction}
An evolution in time of a stream of particles can in general be described
by the Boltzmann equation. In this paper we are interested in
 an evolution of a relativistic gas of massless particles .
  The relativistic Boltzmann equation  is the standard tool in a
study of the evolution of the photon in a gas of charged particles
\cite{rybicki}\cite{dodelson}. A diffusion approximation to the
non-relativistic Boltzmann equation has been applied in many
models of the transport phenomena \cite{landau}. The diffusion
approximation to the relativistic Boltzmann equation is applied in
the heavy ion collisions and the theory of quark-gluon plasma
\cite{ion}\cite{van}. A long time ago Kompaneetz \cite{komp}( see
also \cite{wymann}) derived a non-linear diffusion equation
applicable to the photon propagation in an environment of an
electron gas. The equation describes an evolution of a beam of
photons disturbed by Compton scattering and Bremsstrahlung. These
are the processes widely studied in astrophysics
\cite{rybicki}\cite{ens}\cite{itoh}.

There seems to be the unique way to approximate the
non-relativistic transport phenomena by the diffusion. The concept
of a relativistic diffusion did not achieve such a generally
accepted consensus. Various versions of the relativistic diffusion
are discussed in
\cite{schay}\cite{dudley}\cite{hakim}\cite{deb}\cite{dunkel}\cite{dun}\cite{hor1}\cite{lejan}(for
a review and further references see \cite{talkner}\cite{deb2}). In
refs.\cite{schay}\cite{dudley} the relativistic Brownian motion is
uniquely defined by the requirement that this is the diffusion
whose four momentum stays on the mass-shell and the generator of
the diffusion is the second order relativistic invariant
differential operator. Starting from this basic assumption  we
discussed in \cite{haba} friction terms leading to an equilibrium.
In \cite{habajpa} we studied a diffusion of elementary particles
with spin.

In this paper we show  (sec.3) that if the Kompaneetz equation is
expanded around its equilibrium  ( Bose-Einstein) solution and
non-linear correction terms are neglected then the resulting
diffusion equation is the same as the relativistic diffusion with
friction . First, in sec.2 we formulate the relativistic diffusion
equation for massless particles in a form appropriate for a
comparison with the Kompaneetz equation. In sec.4 we show that the
linearized Kompaneetz equation can be solved applying methods of
quantum mechanics because the relativistic diffusion can
equivalently be described as quantum mechanical evolution in an
imaginary time. Finally, in sec.5 we suggest a generalization of
the Kompaneetz equation to a moving frame and a background
gravitational field. This is a non-linear equation taking into
account the quantum statistics of bosons whose linear
approximation coincides with the (linear) relativistic diffusion
discussed in \cite{lejan}(without friction) and in \cite{habacqg}
(with friction).
\section{The relativistic diffusion of massless particles}
 Following Schay\cite{schay} and
Dudley\cite{dudley} we define the generator $\triangle_{m}$ of the
relativistic diffusion as the $O(3,1)$ invariant second order differential operator on
the mass-shell ${\cal H}_{m}$
\begin{equation}
p^{2}=p_{0}^{2}-p_{1}^{2}-p_{2}^{2}-p_{3}^{2}=m^{2}c^{2}.
\end{equation}
The limit $m\rightarrow 0$ of $\triangle_{m}$ is divergent. In order
to obtain a finite result we  take the limit $m\rightarrow 0$ of
$m^{2}\triangle_{m}$. This limit coincides with another definition
of the diffusion \cite{habajpa}
 as the process generated by $M_{\mu\nu}M^{\mu\nu}$ where $M_{\mu\nu}$
 are generators of the Lorentz group (the
formula for a diffusion of a massless particle has been also
derived in \cite{sorkin} as a continuum limit of   discrete
dynamics of space-time, see also \cite{mex}).

Choosing ${\bf p}$ (boldface type letters denote three dimensional
vectors)
 as coordinates on ${\cal H}_{0}$
we obtain as the limit $m\rightarrow 0$ of $m^{2}\triangle_{m}$ the
operator
\begin{equation}
\triangle_{H}=p_{j}p_{k}\partial^{j}\partial^{k}
+3p_{k}\partial^{k}
\end{equation}
here $k=1,2,3$ and $\partial^{j}=\frac{\partial}{\partial p_{j}}$.
 The relativistic diffusion
equation in the momentum space (a relativistic analog of the
Brownian motion) reads
\begin{equation}
\partial_{\tau}\phi_{\tau}=\frac{1}{2}\gamma^{2}\triangle_{H}\phi_{\tau}.
\end{equation}$\gamma$ is a diffusion constant.
The photon as a massless particle is a typically relativistic
object. At the same time this
 is a quantum system with an internal angular momentum (spin).  In this paper
we neglect the spin. In \cite{habajpa} we discuss the diffusion of
particles with a spin (the helicity in the massless case). It is
shown in \cite{habajpa} that if $m=0$ then the dissipative part of
the evolution does not depend on the helicity . Hence, the neglect
of spin in this paper is justified.

 We add a drag term
\begin{displaymath}
X=Y_{j}\frac{\partial}{\partial
p_{j}}+\gamma^{2}R_{j}\frac{\partial}{\partial
p_{j}}+p_{\mu}\frac{\partial}{\partial x_{\mu}}
\end{displaymath}
to the diffusion (3). The vector field $R$ defines a perturbation
(friction) of the relativistic dynamical system (at $\gamma=0$)
\begin{equation}
\frac{dp_{j}}{d\tau}=Y_{j}, \end{equation} \begin{equation}
\frac{dx^{\mu}}{d\tau}=p^{\mu} \end{equation} where $\mu=0,1,2,3$
and from eq.(1) (for $m=0$) we have $p^{0}=\vert {\bf p}\vert$.
$\tau$ is an affine time parameter describing relativistic
invariant dynamics. Its relation to the laboratory time $x^{0}$ follows
from the zeroth component of eq.(5). We set $Y=0$ till sec.5 where
the vector $Y$ will describe a geodesic motion on a manifold.

 Let
\begin{equation}
{\cal G}=\frac{1}{2}\gamma^{2}\triangle_{H}+X.
\end{equation}
Now, the diffusion equation reads
\begin{equation}
\partial_{\tau}\phi_{\tau}={\cal G}\phi_{\tau}.\end{equation}

 We are interested in the behavior of the relativistic dynamics at large
 values of $\tau$. In particular, whether the diffusion generated by
${\cal G}$ can have an equilibrium limit. We say that  the
probability distribution $dxd{\bf p}\Phi$ is the invariant measure
for the diffusion process  if the statistical mean values are
$\tau$-independent (there may remain the dependence on
$x^{0}$)\begin{equation} \int dxd{\bf p}\Phi(x,{\bf
p})\phi_{\tau}({\bf p},x)\equiv\int dxd{\bf p}\Phi_{\tau}({\bf
p},x)\phi({\bf p},x)=const.
\end{equation}Here,
the evolution of $\Phi$ is defined by
\begin{equation}
\partial_{\tau}\Phi_{\tau}={\cal G}^{*}\Phi_{\tau}
\end{equation}
and ${\cal G}^{*}$ is the adjoint of ${\cal G}$ in $L^{2}(d{\bf
p}dx)$ (the Hilbert space of square integrable functions;the
Lebesgue measure $dxd{\bf p}$ is not Lorentz invariant, the factor
$\vert {\bf p}\vert^{-1}$ necessary for the Lorentz invariance is
contained in $\Phi$, see \cite{haba}\cite{habacqg}).

The $\tau$-invariance (8) is equivalent to \begin{equation}{\cal
G}^{*}\Phi=0.
\end{equation}
Eq.(10) defines a transport equation in the laboratory time
$x^{0}$.

 We denote by $\Phi_{ER}$ the $x$-independent solution of
eq.(10). Then, $R$ can be expressed in terms of $\Phi_{ER}$
\begin{equation}
R_{j}=\frac{1}{2}p_{j}+\frac{1}{2}p_{j}p_{k}\partial^{k}\ln
\Phi_{ER}.
\end{equation}
We assume that $\Phi_{ER}$ is a function of the energy $p_{0}c$
multiplied by a constant  $\beta$ of the dimension inverse to the
dimension of the energy ($\beta=\frac{1}{kT}$ in the conventional
notation, where $T$ denotes the temperature). In such a case  from
eq.(11)
\begin{equation}R_{j}=\frac{1}{2}p_{j}+\frac{1}{2}\beta
c p_{j}\vert {\bf p}\vert (\ln\Phi_{ER})^{\prime}(c\beta\vert{\bf
p}\vert).
\end{equation}
Now, eq.(7) reads
\begin{equation}\begin{array}{l}
\partial_{\tau}\phi_{\tau}=\frac{1}{2}\gamma^{2}\triangle_{H}\phi_{\tau}
 +\frac{1}{2}\gamma^{2}p_{j}(1+\beta c\vert {\bf p}\vert
(\ln\Phi_{ER})^{\prime})\partial^{j}\phi_{\tau}.
\end{array}\end{equation}
Let us consider  an initial probability distribution of the form
$\Phi=\Phi_{ER}\Psi$. Then, from eq.(9) we obtain its evolution
\begin{equation}
\Phi_{\tau}=\Phi_{ER}\Psi_{\tau}
\end{equation}
as a solution of the equation
\begin{equation}
\begin{array}{l}
\partial_{\tau}\Psi_{\tau}=
\frac{1}{2}\gamma^{2}p_{j}p_{k}\partial^{j}\partial^{k}\Psi_{\tau}
+2\gamma^{2}p_{j}\partial^{j}\Psi_{\tau}+\frac{1}{2}\gamma^{2}p_{j}p_{k}(\partial^{j}
 \ln\Phi_{ER})
\partial^{k}\Psi_{\tau}-p_{\mu}\partial_{x}^{\mu}\Psi_{\tau}
\end{array}\end{equation}
(derivatives over the space-time coordinates  have an index
$x$,till sec.5 we restrict ourselves to the diffusion in momentum
space, hence we drop the spatial derivatives). The diffusion
generator (2) is degenerate. This is easy to see if we express it
in the spherical coordinates on ${\cal H}_{0}$
\begin{equation}
p_{0}=r
\end{equation}
$ p_{1}=r\cos\phi\sin\theta $, $p_{2}=r\sin\phi\sin\theta
$,$p_{3}=r\cos\theta$. In these
coordinates\begin{equation}\begin{array}{l}
\triangle_{H}=r^{2}\frac{\partial^{2}}{\partial
r^{2}}+3r\frac{\partial}{\partial r}=\frac{\partial^{2}}{\partial
u^{2}}+2\frac{\partial}{\partial u}\end{array}
\end{equation}
where we  introduced an exponential parametrization of $r$
\begin{equation}
r=\exp u
\end{equation}
Here, $u$ varies over the whole real axis. It follows that the
three dimensional diffusion of massive particles becomes one
dimensional in the limit $m\rightarrow 0$.

In the coordinates (18) it is sufficient if we restrict ourselves to
the drags
\begin{equation}
X=\gamma^{2}R\frac{\partial}{\partial u} \end{equation} The
diffusion (7) is generated by
\begin{equation}
{\cal G}=\frac{\gamma^{2}}{2}(\partial_{u}^{2}+2\partial_{u})
+\gamma^{2}R\partial_{u}.
\end{equation}

  We may write for integrals of spherically symmetric
functions
\begin{equation}
d{\bf p}=4\pi du\exp(3u) .\end{equation} Then \begin{equation}
{\cal G}^{*}=e^{-3u}{\cal G}^{+}e^{3u},\end{equation}where
\begin{equation}\begin{array}{l}
{\cal G}^{+}=\frac{\gamma^{2}}{2}(\partial_{u}^{2}-2\partial_{u})
-\gamma^{2}\partial_{u}R
\end{array}\end{equation}
is the adjoint of ${\cal G}$ in $L^{2}(du)$. The probability
distribution evolves according to eq.(9). The invariant measure
$dxd{\bf p}\Phi_{R}\equiv dxdu\Phi_{I}$  solves an analog of eq.(10)
\begin{equation} {\cal G}^{*}\Phi_{R}= {\cal
G}^{+}\Phi_{I}=0\end{equation}
where
\begin{equation}
\Phi_{I}=\exp(3u)\Phi_{R}.\end{equation}

\section{Linearization of the Kompaneetz equation}
In this section we compare the probability distribution (9)
resulting from the relativistic diffusion equation with the
density of photons moving through an electron gas and described by
the Kompaneetz equation. We assume that $\Phi_{ER}$ in
eqs.(14)-(15) is the Bose-Einstein equilibrium distribution. This
is also the equilibrium solution of the Kompaneetz equation. We
show that if the density of photons is low (so that the quantum
bunching effects can be neglected) then the relativistic diffusion
equation (9) and the Kompaneetz equation linearized around the
equilibrium coincide. We begin with the relativistic diffusion
(9). The $x^{0}$ independent solution of eq.(24) is denoted
$\Phi_{EI}$. The drift $R$ is related to $\Phi_{EI}$ (it is
defined by $\Phi_{ER}$ in eq.(12))
\begin{equation}
R=\frac{1}{2}(\partial_{u}\ln\Phi_{EI}-2).
\end{equation}If the diffusion (9) starts from an initial
distribution
\begin{equation}
\Phi_{I}=\Phi_{EI}\Psi^{I}
\end{equation}
then  using the  relation (26), we obtain the evolution of $\Psi$
\begin{equation}\begin{array}{l}
\partial_{\tau}\Psi^{I}_{\tau}=\frac{1}{2}\gamma^{2}\partial_{u}^{2}\Psi^{I}_{\tau}
+
\frac{1}{2}\gamma^{2}(\partial_{u}\ln\Phi_{EI})\partial_{u}\Psi^{I}_{\tau}.
\end{array}\end{equation}

First, let us consider J\"uttner equilibrium distribution
\cite{jut} in $L^{2}(du)$ (the classical limit of the
Bose-Einstein distribution)
\begin{equation}
\Phi_{EI}^{J}=r^{3}\exp(-\beta c r)\equiv r^{3}n_{J}.
\end{equation}
Then, from eq.(26)
\begin{equation}
R=-\frac{1}{2}(-1+\beta c\exp (u))
\end{equation}The diffusion generator corresponding to the
J\"uttner distribution reads
\begin{equation}\begin{array}{l}
{\cal
G}=\frac{1}{2}\gamma^{2}\Big(\partial_{u}^{2}+3\partial_{u}-\beta
c\exp( u) \partial_{u}\Big)\cr =
\frac{1}{2}\gamma^{2}\Big(r^{2}\partial_{r}^{2}+4r\partial_{r}-\beta
cr^{2}
\partial_{r}\Big).\end{array}\end{equation}
For  the Bose-Einstein distribution (in Compton scattering the
photon number is preserved; if an equilibrium is achieved, then
the chemical potential $\mu\neq 0$, see \cite{wymann})
\begin{equation} \Phi_{EI}^{B}=r^{3}\Big(\exp(\beta (\mu
+cr))-1\Big)^{-1}\equiv r^{3}n_{E}(\mu)
\end{equation}
we have \begin{equation} R=-\frac{1}{2}\gamma^{2}\Big(-1+\beta
c\exp (u)\Big(1-\exp(-\beta\mu-\beta c\exp
u)\Big)^{-1}\Big).\end{equation} Eq.(28) for an
 evolution of $\Psi^{I}$ (with the drift (33)) in the $r$-coordinates reads
\begin{equation}
\begin{array}{l}
\partial_{\tau}\Psi^{I}=\frac{1}{2}\gamma^{2}r^{-2}\Big(\partial_{r}r^{4}\partial_{r}\Psi^{I}
+r^{4}(\partial_{r}\ln
n_{E})\partial_{r}\Psi^{I}\Big),\end{array}\end{equation} where
\begin{equation}
\partial_{r}\ln n_{E}=-\beta c(1-\exp(-\beta\mu-\beta cr))^{-1}.
\end{equation}
In the  high energy limit  we have in the diffusion equation (34)
\begin{displaymath}
\partial_{r}\ln n_{E}\rightarrow -\beta c
\end{displaymath}
(the same as for the J\"uttner distribution).

Now, we can compare the relativistic diffusion equation with the
(non-linear) Kompaneetz equation usually written in the form
\cite{komp}
\begin{equation}
\partial_{\tau}
n=\kappa^{2}\rho^{-2}\partial_{\rho}\Big(\rho^{4}(\partial_{\rho}n
+n+n^{2})\Big).
\end{equation}

 In order to explore the appearance of friction
in eq.(36) we expand the photon distribution of the Kompaneetz
equation around its equilibrium value  in the same way as we
expanded the distribution of the diffusion process (15) around the
equilibrium in eq.(28). If we neglect $n^{2}$ on the rhs of
eq.(36) then the rhs disappears (because
$\partial_{\rho}n_{J}=-n_{J}$, with $\rho=\beta c r$) for the
J\"uttner distribution. Let\begin{displaymath} n=\exp(-\rho)\chi
.\end{displaymath} Then, neglecting the $n^{2}$ term in the
Kompaneetz equation (36) we obtain (after an elementary rescaling
$r\rightarrow \beta c r=x$, $\kappa^{2}=\frac{1}{2}\gamma^{2}$)
the diffusion equation (34) with $n_{E}\rightarrow n_{J}$.

In general, the Kompaneetz distribution $n$ equilibrates to
$n_{E}$ (32)(as $\partial_{\rho}n_{E}+n_{E}+n_{E}^{2}=0$ on the
rhs of eq.(36)). In more detail, let us write the initial
condition for the Kompaneetz equation in the form
\begin{displaymath} n=n_{E}\chi
\end{displaymath}then the evolution of $\chi$ is determined by the
equation
\begin{equation}\begin{array}{l}
\partial_{\tau}\chi_{\tau}=\kappa^{2}\rho^{-2}\Big(\partial_{\rho}\rho^{4}\partial_{\rho}\chi_{\tau}
 +\rho^{4}(\partial_{\rho}\ln n_{E})\partial_{\rho}\chi_{\tau}\Big)
  +\kappa^{2}\rho^{-2}n_{E}^{-1}\partial_{\rho}\Big(\rho^{4}n_{E}^{2}(\chi_{\tau}^{2}-\chi_{\tau})\Big)\end{array}
\end{equation} The last term in eq.(37)  is absent in
eq.(34). It describes the bunching  effect of the quantum
statistics which promotes bosons to condense. As long as the
density of photons is low this  term is negligible.

The Kompaneetz equation finds applications in astrophysics
\cite{rybicki}\cite{peebles}\cite{sun}
\cite{sun2}\cite{sun3}\cite{birk}
 \cite{stebb}\cite{bernstein}\cite{hu}. In particular, the CMBR photons described by
$n_{E}(\mu=0)$ are scattered when passing clusters of galaxies and
other reservoir of hot plasma. When applied to the
Sunyaev-Zeldovitch effect the Kompaneetz equation is usually
discussed only in a linear approximation. In such a case
\begin{equation}\begin{array}{l}\Psi_{\tau} (r)= \int dyy^{-1}P_{\tau}(r,y)\Psi(y)=(2\pi\gamma^{2}\tau)^{-\frac{1}{2}}
\int_{0}^{\infty}
\frac{dy}{y}\Psi(y)\exp\Big(-\frac{1}{2\tau\gamma^{2}}
(\ln\frac{y}{r}-\frac{3}{2}\gamma^{2}\tau )^{2}\Big)
\end{array}\end{equation}
is the solution of the equation
\begin{displaymath}
\begin{array}{l}
\partial_{\tau}\Psi_{\tau}=\frac{1}{2}\gamma^{2}r^{-2}\partial_{r}r^{4}\partial_{r}\Psi_{\tau}
\equiv {\cal G}^{*}\Psi_{\tau}\end{array}\end{displaymath} with
the initial condition $\Psi$. The solution (38) is discussed in
\cite{sun}\cite{bernstein}. For small $\tau$ and the initial
condition $\Psi=n_{E}(\mu=0)$ the solution can be approximated by
\begin{displaymath}
\Psi_{\tau}=n_{E}(\mu=0)+\tau{\cal G}^{*}n_{E}(\mu=0).
\end{displaymath}
 The discussion of Sunyaev-Zeldovich CMBR spectrum distortion
\cite{sun}\cite{birk}\cite{stebb}\cite{bernstein}\cite{carl} is
usually restricted to this approximation.
 The Kompaneetz
equation has been derived from the Boltzmann equation
\cite{komp}\cite{rybicki} under an assumption that the electron
velocities are non-relativistic. The diffusion limit of the
Boltzmann equation with relativistic corrections to the Compton
scattering have been calculated in
\cite{itoh}\cite{stebb}\cite{lesen}.

\section{A relation to the imaginary time quantum mechanics}
In this section we show that the (linear) relativistic diffusion
equation can be solved by means of the methods of quantum
mechanics. We can express solutions of the diffusion equations by
an imaginary time evolution generated by a quantum mechanical
Hamiltonian.The diffusion generator is of the form
\begin{equation}
{\cal G}=\frac{1}{2}\gamma^{2}\partial_{u}^{2}-\omega\partial_{u}
\end{equation}
( we denote a general drift by $\omega$ here in order to
distinguish it from particular $R$ of the earlier sections). Let
\begin{equation}
\Omega(u)=\int^{u}\omega.
\end{equation}
Then
\begin{equation}
\exp(-\Omega){\cal G}\exp(\Omega)\equiv
-H=\frac{1}{2}\gamma^{2}\partial_{u}^{2}-V
\end{equation}
where
\begin{equation}
V=\frac{1}{2}\omega^{2}-\frac{1}{2}\partial_{u}\omega.
\end{equation}

Let
\begin{equation}
\psi=\exp(-\Omega)\phi.
\end{equation}
Then, from eq.(41) it follows
\begin{equation}\begin{array}{l}
\phi_{\tau}(u)=(\exp(\tau {\cal
G})\phi)(u)=\exp(\Omega(u))(\exp(-\tau H)\psi)(u).\end{array}
\end{equation}
 Inserting $\phi=1$ in eq.(44) we obtain $\exp(-\tau
H)\exp(-\Omega)=\exp(-\Omega)$. Hence,
\begin{equation}
H\exp(-\Omega)=0.
\end{equation}

For the J\"uttner model when
\begin{displaymath}
\omega=-\frac{3}{2}\gamma^{2}+\frac{1}{2}\beta c\gamma^{2}\exp u
\end{displaymath}
eq.(42) gives
\begin{equation}
V=\frac{9}{8}\gamma^{4}+\frac{1}{8}\gamma^{4}\beta^{2}c^{2}\exp(2u)-
\frac{1}{4}(1+3\gamma^{2})\gamma^{2}\beta c\exp(u).
\end{equation}
This is a soluble model (the Morse potential ). The eigenvalue
equation for $H$ has a solution in terms of hypergeometric
functions \cite{flugge}.  The propagation kernel determining the
time evolution (44) is also known. The eigenstate of $H$ with the
zero eigenvalue is
\begin{equation}
\exp(-\Omega)=\exp\Big(\frac{3}{2} u-\frac{1}{2}\beta
c\exp(u)\Big).
\end{equation}
The probability density in the state (47)\begin{equation}
\exp(-2\Omega(u))=r^{3}\exp(-\beta c r)=\Phi^{J}_{EI}
\end{equation}
is just the J\"uttner distribution.

 For
Bose-Einstein equilibrium distribution we have $\omega=
-\gamma^{2}-\gamma^{2}R$ where $R$ is defined in eq.(33).
 It can be seen from eq.(45)
that in general \begin{displaymath} \exp(-2\Omega)=\Phi_{EI}
\end{displaymath}
for any equilibrium distribution $\Phi_{EI}$.
\section{Outlook}
There is the unique generalization of the relativistic diffusion
equation (3) to an arbitrary  metric \cite{lejan} describing a
diffusion in the presence of gravity (some other generalizations
are discussed in \cite{deb3}\cite{deb4}). This is the diffusion
which preserves the mass shell
$g^{\mu\nu}p_{\mu}p_{\nu}=m^{2}c^{2}$ ($g^{\mu\nu}$ denotes the
metric tensor). In \cite{habacqg} we have shown that an addition
of a friction is necessary if the particle energy is to be bounded
in time (then it equilibrates). The diffusion equation (3)
(without friction) is explicitly Lorentz invariant as follows from
eq.(6). However, the equilibrium measure cannot be Lorentz
invariant if it is to be normalizable (in a finite volume) and if
the mass shell condition (1) is to be satisfied (there is only one
invariant measure on the mass shell, this is $d{\bf
p}p_{0}^{-1}$). The equilibration to the J\"uttner or
Bose-Einstein distribution takes place in a preferred Lorentz
frame \cite{lorentz}\cite{wel}. In a general frame described by
the four-velocity $w^{\nu}$ we should write the equilibrium
distribution in the form
\begin{equation}
n_{E}(w,\mu)=\Big(\exp(\beta w^{\nu}p_{\nu}+\beta
w^{\nu}\mu_{\nu})-1\Big)^{-1}
\end{equation}
where (by convention) in the rest frame $w=(1,0,0,0)$ and
$\mu_{0}=\mu$ ($\mu_{k}=0$ for $k=1,2,3$). In general, all the parameters ($\beta,\mu,w$) can
depend on the position $x$ (see \cite{mac} for such a framework in
the description of the  quark-gluon plasma).

There is the unique way to generalize the Kompaneetz equation for
a photon distribution function $N(x,p)$ to an arbitrary frame on
the pseudoriemannian manifold in such a way that in the rest frame
on Minkowski space-time its linearized version coincides with the
relativistic diffusion (9). The generalization of eq.(36) reads
\begin{equation}\begin{array}{l}
D_{\tau}N-g^{\alpha\nu}p_{\alpha}\partial_{\nu}^{x}N=\frac{\gamma^{2}}{2}\Big(\triangle_{H}N
\cr+2iw_{\nu}L^{\nu\rho}p_{\rho}N(N+1)+2w^{\nu}p_{\nu}N(N+1)\Big)
\end{array}\end{equation}
where\begin{equation}
D_{\tau}=\partial_{\tau}-Y=\partial_{\tau}-\Gamma_{\alpha\sigma}^{j}p^{\alpha}p^{\sigma}\partial_{j}.
\end{equation}
Here, $\Gamma$ are the Christoffel symbols and $D$ is the
covariant derivative on the cotangent bundle describing the
geodesic motion. $L_{\mu\nu}(x)$ are the generators of the Lorentz
group at the point $x$ on the manifold and $p$ are the coordinates
of the cotangent bundle (the momenta, see \cite{habacqg})
\begin{equation} L_{jk}=-i(g_{jl}p_{k}\partial^{l}-
g_{kl}p_{j}\partial^{l})
\end{equation}
and
\begin{equation}
 L_{0j}=-ig_{jl}\vert {\bf p}\vert
\partial^{l}
\end{equation}
 The lhs of eq.(50) describes the well-known collisionless Boltzmann equation \cite{stewart}.
 The operator $\triangle_{H}$ (2) is the same as the one in the relativistic diffusion equation (3).
 It does not depend on the metric
 (it is invariant under diffeomorphisms of the momenta).
 In eq.(50) we suggest that the collision term resulting
 from the Compton scattering (which could be  calculated
  on the curved manifold following  the elementary derivation in \cite{rybicki})
 has a trivial metric dependence. In the rest frame on the Minkowski
 space-time eq.(50)
coincides with eq.(36). In an arbitrary frame $N_{\tau}$ has the
Bose-Einstein distribution $n_{E}(w,\mu)$ as the equilibrium
measure.

In general, in astrophysics we have moving reference frames and
non-zero gravitational fields. The effects of gravity  seem
to show no observational consequences unless the photons move in an
electron gas around large compact massive objects. We have been
concerned with the evolution  at a large time in this paper. In
astrophysical applications the behavior of the diffusion at large
time (apart from its final effect: the equilibration) has not been
discussed. However, with the increasing sophistication of the
astrophysical observations the gravitational attraction as well as
the large time effects of Compton scattering may give important
information about the sources of the the CMBR distortions and the
spectrum of light coming from compact stars.


\begin{thebibliography}{99}


\bibitem{rybicki}G.B. Rybicki and A.P. Lightman,

Radiative Processes in Astrophysics,Wiley-VCH,1979

\bibitem{dodelson}S. Dodelson, Modern Cosmology, Academic Press, New
York,2003
\bibitem{landau}E.M. Lifshits and L.P. Pitaevskii,


Physical Kinetics, Pergamon Press, Oxford, 1981
\bibitem{ion}B. Svetitsky, Phys.Rev.{\bf D37},2484(1988)

\bibitem{van} H. Van Hess, V. Greco and R.Rapp, Phys.Rev.{\bf
C73},034913(2006)
\bibitem{komp}A.S. Kompaneetz, JETP,{\bf 47},1939(1956) (in
Russian)
\bibitem{wymann}R. Weymann,Phys.Fluids {\bf 8},2112(1965)


\bibitem{ens} T.A. Ensslin and C.R. Kaiser,

Astron.Astrophys.{\bf 360},417(2000)
\bibitem{itoh} N. Itoh, Y. Kohyama and S. Nozawa,

Astrophys.J.{\bf 502},7(1998)
\bibitem{schay} G. Schay, PhD
thesis, Princeton University, 1961
\bibitem{dudley} R. Dudley, Arkiv for Matematik,{\bf 6},241(1965)


\bibitem{hakim}R. Hakim, Journ.Math.Phys.{\bf 9},1805(1968);


\bibitem{deb}F. Debbasch and J.P. Rivet,

Journ.Stat.Phys.{\bf 90},1179(1998)
\bibitem{dunkel}J. Dunkel and P. H\"anggi, Phys.Rev.{\bf E72},036106(2005)

\bibitem{dun}J. Dunkel, P. Talkner and P. H\"anggi,

Phys.Rev.{\bf D75},043001(2007)
\bibitem{hor1}
O. Oron and L.P. Horwitz, Found.Phys.{\bf 35},1181(2005)
\bibitem{lejan}J. Franchi and
Y. Le Jan,

Comm.Pure Appl.Math.{\bf 60},187(2007)

\bibitem{talkner}

J. Dunkel and P. H\"anggi,Phys.Rep.{\bf 471},1(2009)
\bibitem{deb2}C. Chevalier and F. Debbasch,

Journ.Math.Phys.{\bf 49},043303(2008)

\bibitem{haba}Z. Haba, Phys.Rev.{\bf E79},021128(2009)
\bibitem{habajpa}Z. Haba, Journ.Phys.{\bf A42},445401(2009)

\bibitem{habacqg}Z.Haba, arXiv:0909.2880
\bibitem{sorkin}L. Philpott, F. Dowker and R.D. Sorkin,
arXiv:0810.5591
\bibitem{mex}G. Chacon-Acosta and G.M. Kremer, Phys.Rev.{\bf
E76},021201(2007)



\bibitem{jut}F. J\"uttner, Ann.Phys.(Leipzig){\bf 34},856(1911)




\bibitem{peebles}P.J.E. Peebles, Physical Cosmology, Princeton
University Press,1971

\bibitem{sun}R.A. Sunyaev and Ya.B. Zeldovich,
 Astroph.Space.Sci.{\bf
7},20(1970)

\bibitem{sun2}R. Sunyaev and L.G. Titarchuk,
 Astron.Astrophys. {\bf
86},121(1980)



\bibitem{sun3}Y.E. Lyubarsky and R.A. Sunyaev,
 Astron.Astrophys.{\bf
123},171(1983)


\bibitem{birk}M.Birkshaw,Phys.Rep.{\bf 310},97(1999)
\bibitem{stebb}A. Stebbins, arXiv:astro-ph/9705178

\bibitem{bernstein}J. Bernstein and S. Dodelson, Phys.Rev.{\bf D41},354(1990)


 \bibitem{hu}W. Hu, D. Scott and J. Silk, Phys.Rev.{\bf D49},648(1994)


\bibitem{carl}J.E.Carlstrom, G.P. Holder and E.D. Reese,

Ann.Rev.Astr.Astrophys.{\bf 40},643(2002)
\bibitem{lesen}A. Challinor and A. Lasenby, Astroph.J.{\bf
499},1(1998)



\bibitem{flugge}S. Fl\"ugge, Practical Quantum Mechanics,Springer,
Berlin,1974
\bibitem{deb3}F. Debbasch, Journ.Math.Phys.{\bf 45},2744(2004)
\bibitem{deb4} C. Chevalier and F. Debbasch,

Journ.Math.Phys.{\bf 48},023304(2007)

\bibitem{lorentz}J.H. Eberly and A.Kujawski, Phys.Rev.{\bf
155},10(1967)

\bibitem{wel}H.A. Weldon, Phys.Rev.{\bf D26},1394(1982)








\bibitem{mac}T. Matsui,B. Svetitsky and L.D. McLerran,
Phys.Rev.{\bf D34},783(1986)
\bibitem{stewart}J.M. Stewart, Non-equilibrium Relativistic
Kinetic Theory,Lect.Notes in Physics,Vol.10,Springer,1971

\end{thebibliography}
\end{document}